\documentclass[multphys,vecphys]{svmult}

\usepackage{makeidx} 
\usepackage{graphicx}  
\usepackage{multicol}  
\usepackage[latin1]{inputenc}
\usepackage{rotating}

\begin{document}

\title*{Thermonuclear Supernova Explosions and Their Remnants: The case of Tycho}

\author{Carles Badenes\inst{1}\and
Eduardo Bravo\inst{1}\and
Kazimierz J. Borkowski\inst{2}}

\institute{Institut d'Estudis Espacials de Catalunya, Gran Capit\'a 2-4, 08034 Barcelona, Spain
; and Departament de F\'isica i Enginyeria Nuclear, Universitat Polit\'ecnica de Catalunya, 
Av. Diagonal 647, 08028 Barcelona, Spain 
\texttt{badenes@ieec.fcr.es, eduardo.bravo@upc.es}
\and Department of Physics, North Carolina State University, Raleigh NC 27695 
\texttt{kborkow@unity.ncsu.edu}}

\titlerunning{Remnants of Type Ia SNe}

\maketitle

\begin{abstract}
We propose to use the thermal X-ray emission from young supernova remnants (SNRs) originated
in Type Ia supernovae (SNe) to extract relevant information  concerning the explosion mechanism.
We focus on the differences between numerical 1D and 3D explosion calculations, and
the impact that these differences could have on young SNRs. We use the remnant of the Tycho supernova
(SN 1572) as a test case to compare with our predictions, discussing the observational
features that allow to accept or discard a given model.
\end{abstract}

\section{From supernova to supernova remnant}
\label{sec:1}
Thermonuclear supernovae play a key role in our understanding of the origin of elements, 
the chemical evolution of galaxies and the large scale structure of the universe. 
Yet, the many attempts to constrain their progenitor systems and elucidate the detailed 
physics of the explosion mechanism have been unsuccessful so far (see \cite{BK95,HN00} 
for reviews). In \cite{BBBD03} 
we established a connection between the thermal X-ray emission from shocked ejecta in the young 
supernova remnants originated by Type Ia explosions and a grid of 1D theoretical 
supernova models, including examples from all the explosion mechanisms currently under debate: 
pure deflagrations, delayed detonations, pulsating delayed detonations 
and sub-Chandrasekhar explosions. 
We showed that the different density and chemical composition profiles
for the ejecta predicted by the explosion models have a profound impact on the 
dynamics of the young SNRs and result in different ionization states and electron temperatures for the 
shocked plasma, and therefore different emitted thermal X-ray spectra for each model. 
Significant conclusions about the explosion models can be drawn using this technique; in particular,
the differences between 1D and 3D explosion models are large enough to have a noticeable impact on the 
spectrum of the shocked ejecta several hundred years after the explosion itself. 

\section{Discussion of the analysis technique}
\label{sec:2}
In \cite{BBBD03} an adiabatic 1D hydrodynamic code was used to follow the evolution of the SNR models.
The use of an adiabatic code is justified by the absence of
optical emission from radiatively cooled knots in the well known Type Ia
SNRs Tycho and SN1006, in contrast to some core-collapse SNRs like Cas A, 
which do have such emission. We have calculated the order of magnitude of
the radiative and ionization losses in the ejecta, and found that, except for a few extreme cases, 
their impact on the
dynamics was negligible up to several thousand years after the explosion. 
Other factors that could compromise this simple modeling 
technique, like thermal conduction, would require even
longer timescales, of the order of 10,000 yr, to have an impact on the overall evolution. 
Deviations from 1D dynamics, either produced during the formation
of the remnant or already present in the ejecta or ambient medium (AM), are difficult to quantify \cite{WC01}. 
Type Ia SNRs seem to be more spherically symmetric than core-collapse SNRs, with significantly 
less turbulent dynamics, but the amount 
of clumping present in Type Ia SN ejecta is not known, even if observations seem to imply that
it is not very large (see Bravo \& Garc\'ia-Senz, these proceedings, and references therein). 
The presence of clumps in the shocked ejecta will not affect our conclusions unless 
their density contrast is very large. It is worth noting that
analysis of some of the bright ejecta 
knots in Cas A seems to suggest that they have a low density contrast \cite{LH03},
but this remains an open question, and needs to be adressed with 3D models and high resolution 
X-ray observations.
 
\section{Explosion models: 3D vs 1D}
\label{sec:3}
Numerical calculations of thermonuclear supernova explosions in one dimension have become a commonplace 
benchmark for the analysis
of Type Ia SNe, but their validity is questionable because the subsonic combustion 
fronts (deflagrations) that play a fundamental role in all of them are subject to 
instabilities, and therefore cannot be simulated with 1D codes 
in a self-consistent way.

In recent times, the first three dimensional (3D) calculations of Type Ia explosions
have begun to appear in the literature: see \cite{G03},\cite{R02} and the paper by Bravo \& Garc\'ia-Senz
in these proceedings. A common feature of
all these calculations, and the most remarkable difference between 1D and 3D models, is the 
uniform mixing of unburnt C and O material with \(^{56}\)Ni and other elements
throughout the ejecta. 
This mixing should have an impact on the optical spectra of the supernovae and on the thermal 
X-ray spectra from the shocked ejecta in the SNRs, 
but neither of these signatures has been confirmed so far.
\cite{G03} pointed out that no evidence for low-velocity C and O was found in optical spectra 
of Type Ia SNe, but this assertion is being revised \cite{BLH03}. Spatially resolved spectroscopy
of Type Ia SNRs also provides indirect evidence for some kind of composition 
stratification in the shocked ejecta \cite{HG97,D01,L03}.
The absorbed UV spectrum of the Schweizer-Middleditch star, which is placed behind the remnant 
of SN1006, poses an alternative observational constraint for the presence of mixing in
Type Ia SN ejecta. No evidence has been found for C or O absorption lines in HST observations of this star \cite{H97},
implying that these elements, if present, would have to be in 
ionization states very different from those of the observed Si and Fe.

\section{Spectral signatures of C and O mixing in the ejecta of young SNRs}
\label{sec:4}
\begin{figure}
\centering
\includegraphics[height=8cm,angle=90]{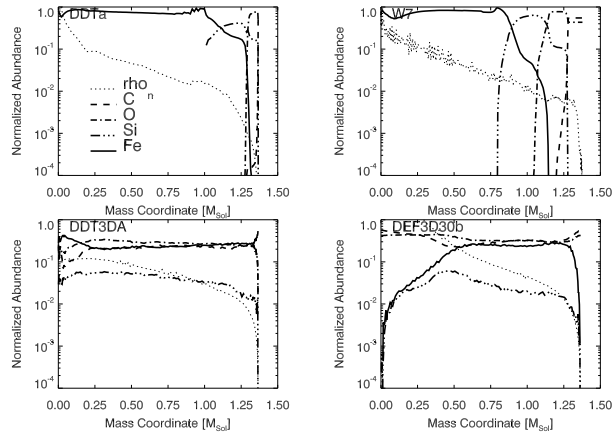}
\caption{Composition and normalized density profiles of the four Type Ia models discussed in
this paper: DDTa, W7, DDT3DA and DEF3D30b. For clarity, only C, O, Si and Fe are represented.
The thin dotted line is the normalized density profile}
\label{fig:1}    
\end{figure}

In Fig.\ref{fig:1} we present the composition and density profiles of four Type Ia
explosion models. DDTa is a 1D delayed detonation \cite{BBBD03}; W7 is a 1D 'fast' 
deflagration \cite{NTY84}; DDT3DA and DEF3D30b are 1D mappings of 3D models 
described by Bravo \& Garc\'ia-Senz elsewhere in these proceedings: DDT3DA is the delayed 
detonation with the transition where \textit{D}>2.5 and DEF3D30b is the deflagration with 30 equal size bubbles.
The mixing of unburnt C and O with \(^{56}\)Ni (which we have represented as Fe, the product of its decay) 
is apparent in the two 3D models,
in contrast with the layered structure of the 1D models.

One of the main conclusions of \cite{BBBD03} was that the density enhancement effect towards
the contact discontinuity between ejecta and ISM, first pointed out in \cite{DC98} and confirmed by 2D
simulations in \cite{D00}, dominates the integrated thermal X-ray emission from 
the shocked ejecta. Self-consistent calculation of electron heating reveals that this is also the hottest region.
This leads to important contributions to the spectrum from elements that were not 
synthesized in large quantities in the explosion, but are located in hot, high density
regions in the young SNR (X-ray emissivity scales as \(\rho^2\)). 
The relative smoothness of the chemical composition profile for the 3D explosion 
models, however, tends to assuage this effect, and the relative contribution from each element
to the emitted spectrum will be more closely related to the total amount of the element synthesized in
the explosion.

\begin{figure}
\centering
\includegraphics[height=8cm,angle=90]{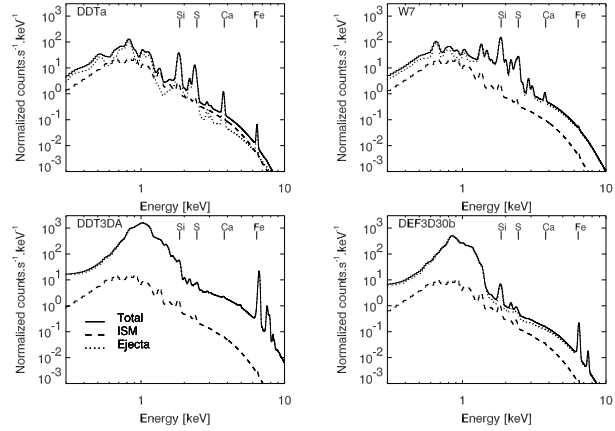}
\caption{Synthetic spectra at the age of Tycho for the four models. The
contributions from shocked ejecta and ISM are shown with dotted and dashed lines, respectively.}
\label{fig:2}    
\end{figure}
\begin{figure}
\centering
\includegraphics[height=4cm,angle=270]{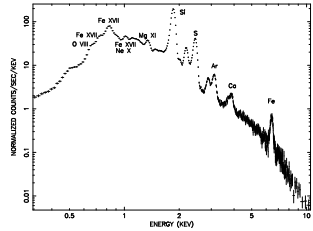}
\caption{XMM-EPIC MOS1 spectrum of the Tycho SNR, from \cite{D01}}
\label{fig:3}      
\end{figure}

In Fig.\ref{fig:2} the total integrated spectrum for the four models at the age of the Tycho SNR is shown, 
calculated following the scheme described in \cite{BBBD03} and convolved with the instrumental response 
of the XMM-Newton Epic-MOS1 camera to facilitate comparison with the observed spectrum from the Tycho SNR, 
which is given in Fig.\ref{fig:3}. No Ar data are present
in our atomic code, but there is enough information in the
K\(\alpha\) lines of Si, S, Ca and Fe, marked in Fig.\ref{fig:2}, and in the overall spectral shape, to extract 
significant conclusions.
The large amount of Fe in the outer layers of the 3D models, which turn out to be the densest and hottest regions
in the shocked ejecta throughout the evolution of the SNR (see \cite{BBBD03}), 
dominates their emitted spectrum. The strength of the Fe L complex around 1 keV and the Fe K\(\alpha\) 
line at 6.4 keV are incompatible with the observed spectrum of Tycho. 
The Fe emission is so strong, in fact, that the Fe continuum masks the 
contribution from Si, S and Ca, which are clearly seen in the observed spectrum. Another interesting feature of the 3D 
models is the presence of the Ni K\(\alpha\) line at 7.5 keV. This line comes from stable Ni, whose parent nuclei are 
synthesized in small amounts together with \(^{56}\)Ni, and it has not been observed so far in any young 
Type Ia SNR. The large amount of C and O in the ejecta does not have a clear imprint on the spectrum
because lines of these elements are located at low photon energies where the interstellar absorption
and the poor spectral resolution of the XMM-Newton EPIC makes their detection difficult.
The 1D models, on the other hand, show important contributions from Si, S and Ca, albeit 
the overall spectral shape is reproduced much better (but not perfectly!) by 
the DDTa model than by the W7 model. The absence of the
Fe K\(\alpha\) line in the W7 model is due to the fact that the Fe is at smaller radii, and therefore
at a lower density and in a lower ionization state.

We conclude that the X-ray spectrum of the Tycho SNR cannot be explained with the current 3D explosion
models for Type Ia SNe using the simulation scheme described in \cite{BBBD03}. The mixing of burnt and unburnt
material throughout the supernova ejecta seems to contradict a number of independent observations, casting a doubt
on the validity of the current 3D explosion models, at least in their present state. This is a preliminary
conclusion; a final resolution of this issue should be based on more sophisticated
hydrodynamical simulations of Type Ia SNRs.

%%%%%%%%%%%%%%%%%%%%%%%% referenc.tex %%%%%%%%%%%%%%%%%%%%%%%%%%%%%%
% sample references
% "physics"
%
% Use this file as a template for your own input.
%
%%%%%%%%%%%%%%%%%%%%%%%% Springer-Verlag %%%%%%%%%%%%%%%%%%%%%%%%%%

%
%BibTeX users please use
% 

%\bibliographystyle{}
%\bibliography{}
%
% Non-BibTeX users please use

\printindex
\end{document}